\documentclass[sigconf]{acmart}
\AtBeginDocument{%
  \providecommand\BibTeX{{%
    \normalfont B\kern-0.5em{\scshape i\kern-0.25em b}\kern-0.8em\TeX}}}

\usepackage{url}

\newcommand{\multisum}{\texttt{codet5-base-multi-sum}}

\usepackage{tabularx}
\usepackage{multirow}
\acmConference[WORKS 2023]{18th Workshop on Workflows in Support of Large-Scale Science}{November 12 2023}{Denver, CO}
\usepackage{xcolor}
\definecolor{brickred}{RGB}{203, 65, 84}
\definecolor{RedOrange}{RGB}{255, 69, 0}
\definecolor{Mahogany}{RGB}{192, 64, 0}
\definecolor{Maroon}{RGB}{128, 0, 0}
\definecolor{RedViolet}{RGB}{199, 21, 133}
\definecolor{DarkRed}{HTML}{8B0000}
\definecolor{Crimson}{HTML}{DC143C}
\definecolor{FireBrick}{HTML}{B22222}
\usepackage[frozencache,cachedir=minted-cache]{minted}

\begin{document}

\title{\texttt{Laminar}: A New Serverless
Stream-based 
Framework with Semantic Code Search and Code Completion }

\author{Zaynab Zahra}
\affiliation{%
\institution{University of St. Andrews}
  \city{St Andrews}
  \country{UK}}
\email{zz46@st-andrews.ac.uk}

\author{Zihao Li}
\affiliation{%
\institution{University of St. Andrews}
  \city{St Andrews}
  \country{UK}}
\email{zl81@st-andrews.ac.uk}

\author{Rosa Filgueira}
\affiliation{%
\institution{University of St. Andrews}
  \city{St Andrews}
  \country{UK}}
\email{rf208@st-andrews.ac.uk}


\begin{abstract}
  This paper introduces \texttt{Laminar}, a novel serverless framework based on \texttt{dispel4py}, a parallel stream-based dataflow library. \texttt{Laminar} efficiently manages streaming workflows and components through a dedicated registry, offering a seamless serverless experience. Leveraging large lenguage models, \texttt{Laminar} enhances the framework with semantic code search, code summarization, and code completion. This contribution enhances serverless computing by simplifying the execution of streaming computations, managing data streams more efficiently, and offering a valuable tool for both researchers and practitioners.
\end{abstract}


\maketitle

\section{Introduction}
\label{section:intro}
%[Note for ROSA: DRAFT OF INTRODUCTION - TO BE IMPROVED LATER]

In the rapidly evolving landscape of cloud computing, serverless computing~\cite{kumar2019serverless} has emerged as a transformative paradigm, offering scalability, cost-effectiveness, and simplicity in deploying applications. However, the surge in data-intensive applications and the increasing demand for real-time processing present new challenges~\cite{shafiei2022serverless} for existing serverless frameworks. Firstly, traditional serverless architectures struggle to efficiently handle the continuous flow of streaming data, resulting in  bottlenecks and latency issues. Secondly, supporting stateful computations within a serverless environment becomes complex due to the need to maintain and manage the state across distributed and ephemeral instances.

To address these challenges we introduce \texttt{Laminar}~\footnote{\url{https://github.com/dispel4pyserverless}}, a novel open-source Serverless Stream-based Processing Framework with Deep Learning Code Search. Unlike traditional serverless frameworks, \texttt{Laminar} provides a comprehensive solution that seamlessly handles data streams and supports stateful computations by leveraging the power of \texttt{dispel4py} Python library. \texttt{dispel4py} inherent support for parallelism enables concurrent data processing, while abstract workflow descriptions in Python empower users to construct intricate stream processing pipelines. Similar to how functions encapsulate specific tasks in traditional programming, \texttt{dispel4py} Processing Elements (PEs) represent modular computational units within the \texttt{Laminar} serverless environment. Furthermore, \texttt{Laminar} goes beyond existing serverless frameworks by providing novel deep learning search facilities to find relevant PEs.  The main contributions of this work are: 

%\vspace{-\baselineskip}
\begin{itemize} 
\setlength\itemsep{2pt}
\item Handling Data Streams and Stateful Computations: \texttt{Laminar} bridges a gap in conventional serverless frameworks, providing native support for data streams and stateful computations through an intuitive workflow description system and a streamlined registry. %enhancing the execution of data-intensive and real-time processing applications.
\item Endpoints: \texttt{Laminar} implements comprehensive server-client architecture with endpoints to facilitate communication for registering, executing, and managing PEs and workflows.
\item Intuitive Client Functionality: \texttt{Laminar} introduces an intuitive client interface for convenient PE and workflow registration, execution, and management. It automates library detection and passes the information to the server for remote PE execution.

\item Serverless Execution Engine: Leveraging the principles of serverless computing, \texttt{Laminar} efficiently handles PE and workflow execution by automatically provisioning resources and installing the necessary libraries ensuring seamless serverless operation.

\item Registry: \texttt{Laminar} provides a robust functionality for registering workflows, including PEs and their properties. The registry serves as a central repository, enhancing \texttt{Laminar}'s efficiency and management of serverless components. 

\item Deep Learning Code Search Facilities: \texttt{Laminar} harnesses the power of large language models, enhancing its capabilities for advanced PE code search, code summarization, and code completion. This integration significantly enhances semantic code search functionality. We conducted evaluations of multiple models to determine the optimal ones for integration into \texttt{Laminar}.
\end{itemize}

%\texttt{Laminar} represents a significant advancement in serverless computing by combining \texttt{dispel4py}'s power with text-to-code and code-to-code facilities and stream-based processing. As the demand for real-time data processing and data-intensive applications continues to grow, \texttt{Laminar} emerges as a valuable tool simplifying development and execution of workflows, empowering users to harness the full potential of serverless computing and deep learning technologies. 

The remainder of the paper is structured as follows. Section~\ref{section:background} presents background on technologies relevant for this work. Section~\ref{section:laminar} details the features of \texttt{Laminar}. While Section~\ref{section:searches} specifically focuses on the different registry searches supported by \texttt{Laminar}. The practical utility of \texttt{Laminar} is demonstrated through two showcased use cases in Section~\ref{section:use-cases}. Section~\ref{section:evaluation} presents different evaluations performed with this framework. We contextualize our work by surveying related studies in Section~\ref{section:related-work}, and finally, we conclude and outline future directions in Section~\ref{section:conclusions}.

%By extending its functionality, Laminar introduces text-to-code and code-to-code facilities powered by the integration of the Unixcoder Large Language model. This integration empowers Laminar to facilitate seamless text-to-code translation, enabling researchers and practitioners to harness the power of serverless computing while exploring and leveraging a wide range of deep learning algorithms.

\section{Background}
\label{section:background}

In the following sections, we look into the background work that is inherently established in this paper.

\subsection{\texttt{dispel4py}}
\label{section:dispel4py}

dispel4py~\cite{filgueira2016dispel4py, LIANG2022102} is a parallel stream-based dataflow framework designed for creating and executing data-intensive applications. It provides an abstract and user-friendly approach to workflow creation, including automatic parallelization, which allows users to design, execute, and optimize workflows without requiring in-depth knowledge of the underlying hardware or middleware. Key concepts in \texttt{dispel4py} include: 

\begin{itemize}

\item Processing Element (PE): The fundamental unit in \texttt{dispel4py}, acting as a computational task within a workflow graph. PEs connect through inputs and outputs for seamless stream-based data flow. They can be stateful, retaining previous inputs, or stateless, focusing on current data. Various PE types are available for distinct roles: \texttt{ProducerPE} (one output port), \texttt{IterativePE} (one input port, one output port), \texttt{ConsumerPE} (one input port), \texttt{GenericPE} (custom-defined, any number of ports).

\item Instance: Refers to a PE execution within the computing process. Multiple instances of a single PE can be assigned, allowing workflow scaling.

\item Connection: Defines data flow between PEs, determining data consumption and production rates. %To ensure smooth data transfer, connections provide adequate buffering.

\item Mappings: Maps workflows onto execution systems. These include a \textit{Simple} mapping for running workflows sequentially, and parallel options such as \textit{MPI}~\cite{10.5555/898758}, \textit{Redis}~\cite{eddelbuettel2022brief}, and \textit{Multiprocessing} (\textit{Multi})~\footnote{\url{https://docs.python.org/3/library/multiprocessing.html}}, eliminating the need for manual workflow modifications.

\item Abstract Workflow: Represents logical connections between PEs, outlining computational sequences and data transformations. It is what the user describes.

\item Concrete Workflow: During enactment (after the user specifies the mapping to use and the number of processes), \texttt{dispel4py} automatically builds the concrete (parallel~\footnote{All mappings, except \textit{Simple}, trigger the generation of an automatic parallel concrete workflow. This concrete workflow is capable of accommodating multiple parallel patterns, including the \textit{farm} and \textit{pipeline} patterns.}) workflow, which is a directed acyclic graph (DAG) based on the abstract workflow. The concrete workflow is the actual workflow executed by the compute infrastructure.

\item Grouping: Specifies how PEs communicate during input connections. For example, \texttt{group-by} (see Listing~\ref{code:countPE}), which behaves similarly to `MapReduce', routing data units with the same value in the specified element to the same PE instance.
%\item Grouping: Grouping specifies the communication pattern between PEs for an input connection. There are four different groupings available: shuffle, group-by, one-to-all, and all-to-one. Each grouping arranges a set of receiving PE instances. %For example, shuffle randomly distributes data units to the instances, while group-by ensures that each value in the specified elements of each data unit is received by the same instance.

%\item Workflow Execution Time: The workflow execution time is the time required to finish all PE instances in the workflow. In \texttt{dispel4py}, the execution time varies depending on the number of processes used for running the workflow. %Ideally, increasing the resources should result in a lower execution time. 

\end{itemize}

\begin{figure}[htbp]
\centerline{\includegraphics[width=0.48\textwidth]{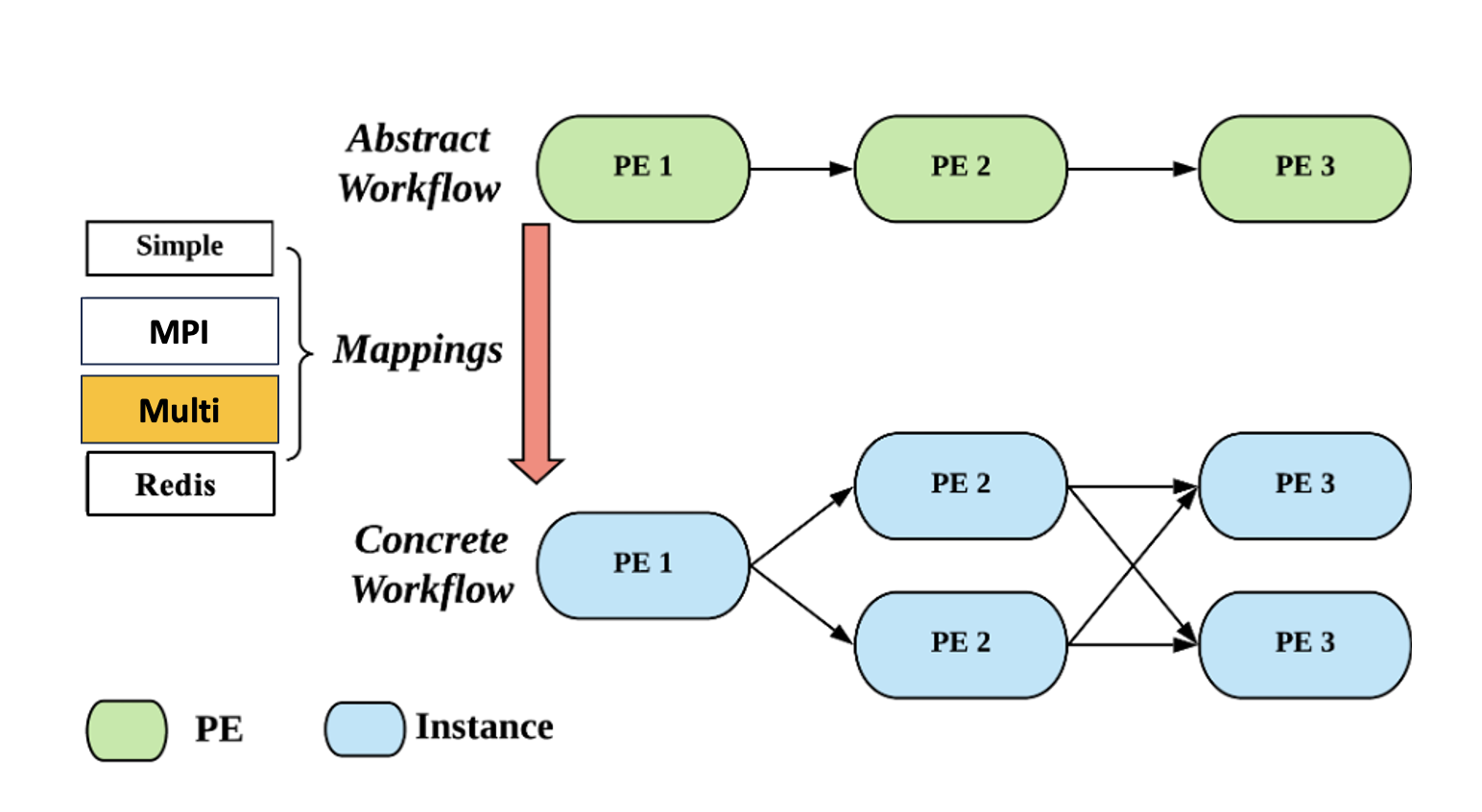}}
\caption{Example of a \texttt{dispel4py} workflow, in which the user has indicated to run it using the \textit{Multi} mapping with five processes. Each PE instance runs in a different process.}
\label{fig:dispel4py}
\end{figure}

To create \texttt{dispel4py} workflows, users implement PEs and connect them in graphs. In Figure~\ref{fig:dispel4py}, an example workflow illustrates three PEs distributed among five processes (e.g., one PE instance for PE1 and two for PE2 to PE3) using the \textit{Multi} mapping. Users design the green abstract workflow, while the blue concrete workflow is automatically generated during enactment. 

Listing~\ref{code:numberproducer} provides the code for the first Processing Element (PE1) in the dispel4py workflow shown in Figure~\ref{fig:dispel4py}. Full workflow code is available in Listing~\ref{listing:workflow-code}. Note that the core functionality of a PE is encapsulated within the \texttt{\_process} function.

\begin{listing}[h]
\begin{minted}[mathescape,
               numbersep=5pt,
               gobble=2,
               frame=single,
               framesep=1mm]{python}
    class NumberProducer(ProducerPE):
        def __init__(self):
            ProducerPE.__init__(self)   
        def _process(self):
            # Generate a random number
            result= random.randint(1, 1000)
            # Return the number as the output
            return result
\end{minted}
\caption{\texttt{NumberProducer} \textit{stateless} PE, referred to as \texttt{PE1} in Figure~\ref{fig:dispel4py}, generates random numbers and streams them out.}
\label{code:numberproducer}
\end{listing}

%Note that \texttt{dispel4py} is a versatile framework capable of handling both stateful and stateless workflows, providing native support for seamless execution of PEs with persistent state or stateless computations.

\vspace{-0.2cm}

\subsection{Serverless Computing}
\label{section:serverless}

Serverless computing~\cite{10.1145/3406011}, also known as Function-as-a-Service (FaaS), is a cloud computing paradigm that abstracts away server management and infrastructure concerns from developers. In a serverless architecture, developers focus solely on writing and deploying individual functions or code snippets to the cloud, and the cloud provider takes care of automatically provisioning, scaling, and managing the underlying infrastructure needed to execute those functions. This approach allows developers to focus on the core logic of their applications, without worrying about server maintenance, resource management, or scalability.

In a traditional cloud computing model, developers typically need to manage virtual machines (VMs) or containers to run their applications. This can be time-consuming and resource-intensive, especially when handling fluctuating workloads. Serverless computing, on the other hand, abstracts away the concept of servers entirely, allowing developers to run their functions on demand without the need to explicitly manage the infrastructure. Functions are executed in stateless containers that are spun up and scaled automatically, based on the incoming workload.

One of the key advantages of serverless computing is its cost-effectiveness. Users are billed only for the actual execution time of their functions, rather than for idle server time. This pay-per-invocation pricing model can result in significant cost savings, especially for applications with variable workloads.

\subsection{Language Models and Transformers}
\label{section:LM}

The landscape of computer capabilities in comprehending human languages, speaking, and reading has undergone a transformation due to the advent of natural language processing models. Among the state-of-the-art models, the transformer architecture has demonstrated remarkable advancements~\cite{DBLP:journals/corr/abs-1910-03771}. However, the scope of these models extends beyond human language and can be expanded to incorporate abstract syntax trees, enabling them to understand and compare code. In this work, we have applied three different language models:
%\vspace{-\baselineskip}
\begin{itemize} 
 \item \textit{UnixCoder}\cite{unixcoder} is a specialized transformer-based model designed to convert Abstract Syntax Trees (ASTs) into sequential text representations~\cite{LOPEZESPEJEL2023100013}. This model enhances the semantic representation of code fragments through embeddings, achieved by employing multi-modal contrastive learning (MCL) for comprehensive code semantic capture using ASTs, and cross-modal generation (CMG) to align embeddings across different programming languages via code comments.  In extensive experiments~\cite{unixcoder}, \textit{UnixCoder} outperformed state-of-the-art models like CodeBERT~\cite{codebert}, GraphCodeBERT~\cite{graphcodebert} and  SYNCOBERT~\cite{syncobert}  in various code-related understanding tasks, including semantic code search  and clone detection. 
 
 \item \textit{ReACC-py-retriever} is a model developed for \textit{ReACC} retrieval-augmented code completion Framework ~\cite{lu2022reacc}. The model utilizes “external” context for the code completion task by retrieving semantically and lexically similar codes from existing codebase. It has been pre-trained with a dual-encoder as a retriever for partial code search, which retrieves code fragments given a partial code. On the CodeXGLUE~\cite{codexglue} benchmark, the model achieves a state-of-the-art performance in the code completion task.

 \item \textit{CodeT5}~\cite{codet5} is a pre-trained encoder-decoder model that incorporates the token type information from code  and allow for multi-task learning on downstream tasks. \textit{CodeT5} is  with three code intelligence capabilities: 1) \textit{text-to-code generation} to generate code based on the natural language description ; 2) \textit{Code autocompletion}, to complete the whole function of code given the target function name; and 3) \textit{Code summarization} to generate the summary of a function in natural language description. Among those tasks, \textit{CodeT5} achieves state-of-the-art performance for code summarization task. 
 
\end{itemize}

\subsection{Semantic Code Search Paradigms}
\label{section:code-search}

\textit{Semantic code search}~\cite{guo2019deep} is a critical functionality in \texttt{Laminar}, enabling \textit{text-to-code} similarity queries over the registry. This NLP technique allows searching for existing
code snippets through natural language, which can greatly improve programming efficiency.  The objective is to identify matching codes in the search corpus that correspond to the query.

\begin{figure}[h!]
\includegraphics[width=\linewidth]
{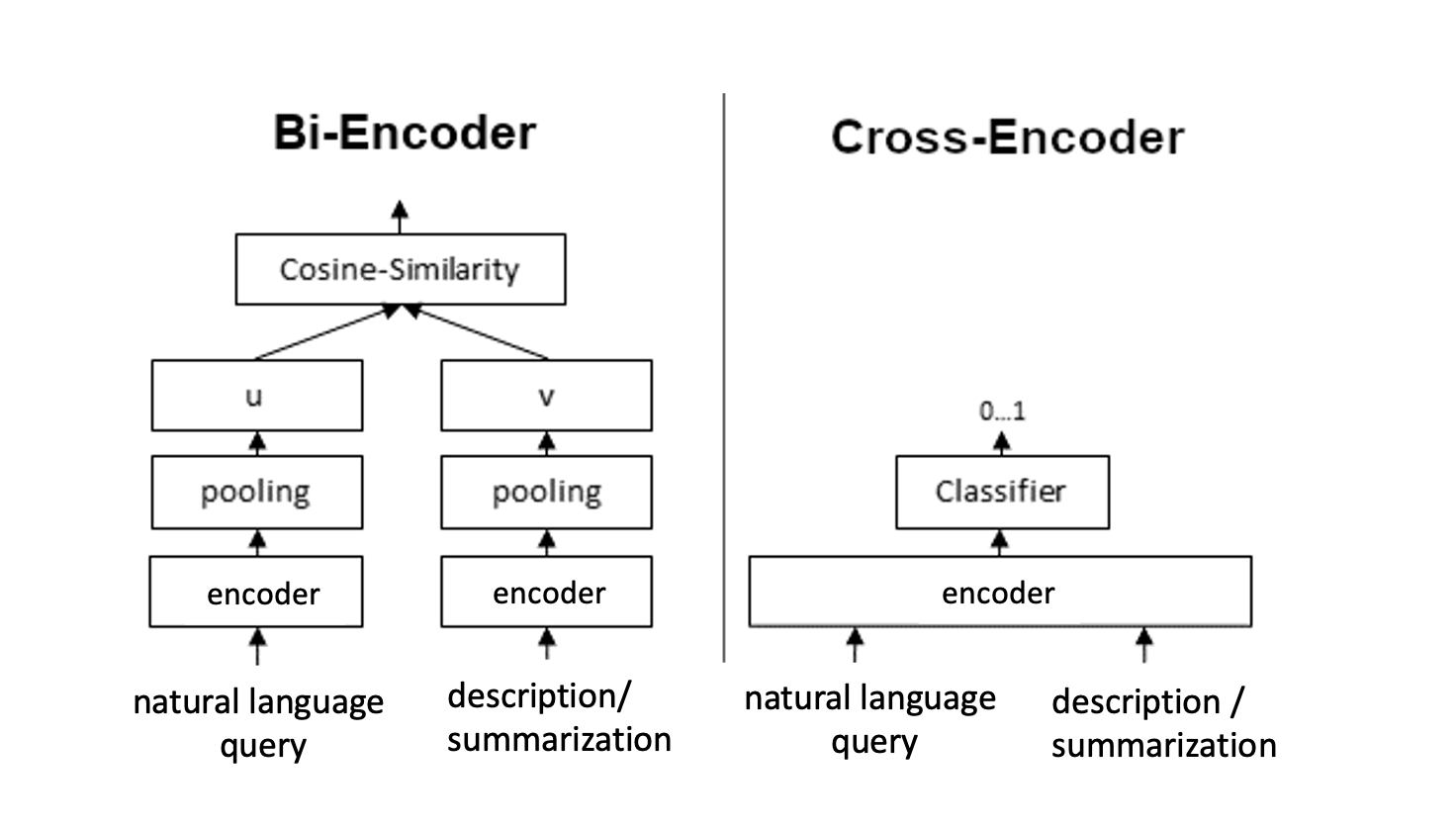}
\vspace{-0.2cm}
\caption{The concept diagram of \textit{bi-encoder} (left) and \textit{cross-
encoder} (right) code search architecture.}
\label{fig:encoders}
\end{figure}

We adopted the bi-encoder paradigm, as illustrated in Figure~\ref{fig:encoders}, where each input (natural language query or a code description/ summarization) is independently mapped into a dense vector space. Bi-encoders calculate embeddings for both inputs, enabling efficient storage of embeddings for subsequent queries. In contrast, cross-encoders perform full-attention over the input pairs of sentences, resulting in better accuracy but reduced efficiency. The bi-encoder architecture in \texttt{Laminar} is well-suited for applications like PEs similarity, where fast querying of separate embeddings is crucial. While bi-encoders are faster, cross-encoders achieve better accuracy but may not be practical in certain scenarios. In summary, the bi-encoder approach strikes a balance between efficiency and effectiveness, making it an ideal choice for searches in \texttt{Laminar}.

\subsection{Code Completion and Code Summarization}
\label{section:code-completion}

In the context of this work, \textit{Code completion}~\cite{wang2019deep} involves predicting subsequent code tokens based on the given code context, contributing to enhanced programming productivity. Recent advancements have demonstrated the efficacy of statistical language modeling with transformers in improving code completion performance by leveraging vast source code datasets.

Conversely, code summarization~\cite{sym14030471} refers to the generation of concise and coherent natural language summaries or descriptions that encapsulate the core functionality and behavior of a specific segment of source code. This task aims to facilitate comprehension and documentation by providing human-readable explanations for intricate code snippets, functions, or methods.

In the \texttt{Laminar} framework, code completion serves the purpose of allowing users to input incomplete PE snippets for completion queries. The framework then retrieves similar PEs from our registry, which functions as our search corpus. Moreover, code summarization is leveraged for storing descriptions of PEs within the registry when users fail to provide them. This dual approach enhances the overall usability and efficiency of the framework's capabilities.

clone-detection, as presented in~\cite{unixcoder}. Although the base model provided by the authors through Hugging 
Face~\footnote{\url{unixcoder-base}} lacks fine-tuning tasks, we undertook our own fine-tuning process following instructions 
outlined 
in~\footnote{\url{https://github.com/microsoft/CodeBERT/tree/master/UniXcoder/downstream-tasks/}}. 
This process involved utilizing the AdvTest dataset~\cite{codexglue}, which comprises 280,634 pairs of (documentation, function) sourced from CodeSearchNet~\cite{codesearchnet}. Notably, the dataset normalizes Python function and variable names to enhance model understanding and generalization capabilities. Our fine-tuning efforts led to the development of two models: \textit{unixcoder-code-search} and \textit{unixcoder-clone-detection}. It is important to highlight that the fine-tuned model was originally developed by the authors of this owrk for another complementary study~\cite{reposim}, which focused on comparing repository similarities. Each of these models underwent approximately 6 hours of fine-tuning on an NVIDIA A40 GPU server..

\section{\texttt{Laminar} Overview}
\label{section:laminar}
This section presents an overview of the fundamental components that constitute the \texttt{Laminar} architecture, each serving a distinct purpose within the framework. The core  elements encompass the \textit{Registry}, \textit{Server}, \textit{Execution Engine}, and \textit{Client}, as detailed in Table~\ref{table:laminar-core}.
\begin{table}[h!]
\centering
\begin{tabular}{ |m{3.5em}|m{17em}|m{4em}|}
\hline
\textbf{Element} & \textbf{Purpose} & \textbf{Section} \\
\hline
Registry & Stores user, PE, and workflow information &Section~\ref{section:registry}.
\\
\hline
Server & Coordinates system functionality & Section~\ref{section:server}. 
\\
\hline

Execution Engine & Enables serverless workflow execution & Section~\ref{section:execution} \\
\hline

Client & Interacts with server and users requests & Section~\ref{section:client} \\
\hline
\end{tabular}
\caption{\texttt{Laminar} Core Elements}\label{table:repo-services}
\label{table:laminar-core}
\vspace{-0.4cm}
\end{table}

%\vspace{-0.4cm}

\begin{figure}[h!]
\includegraphics[width=\linewidth]{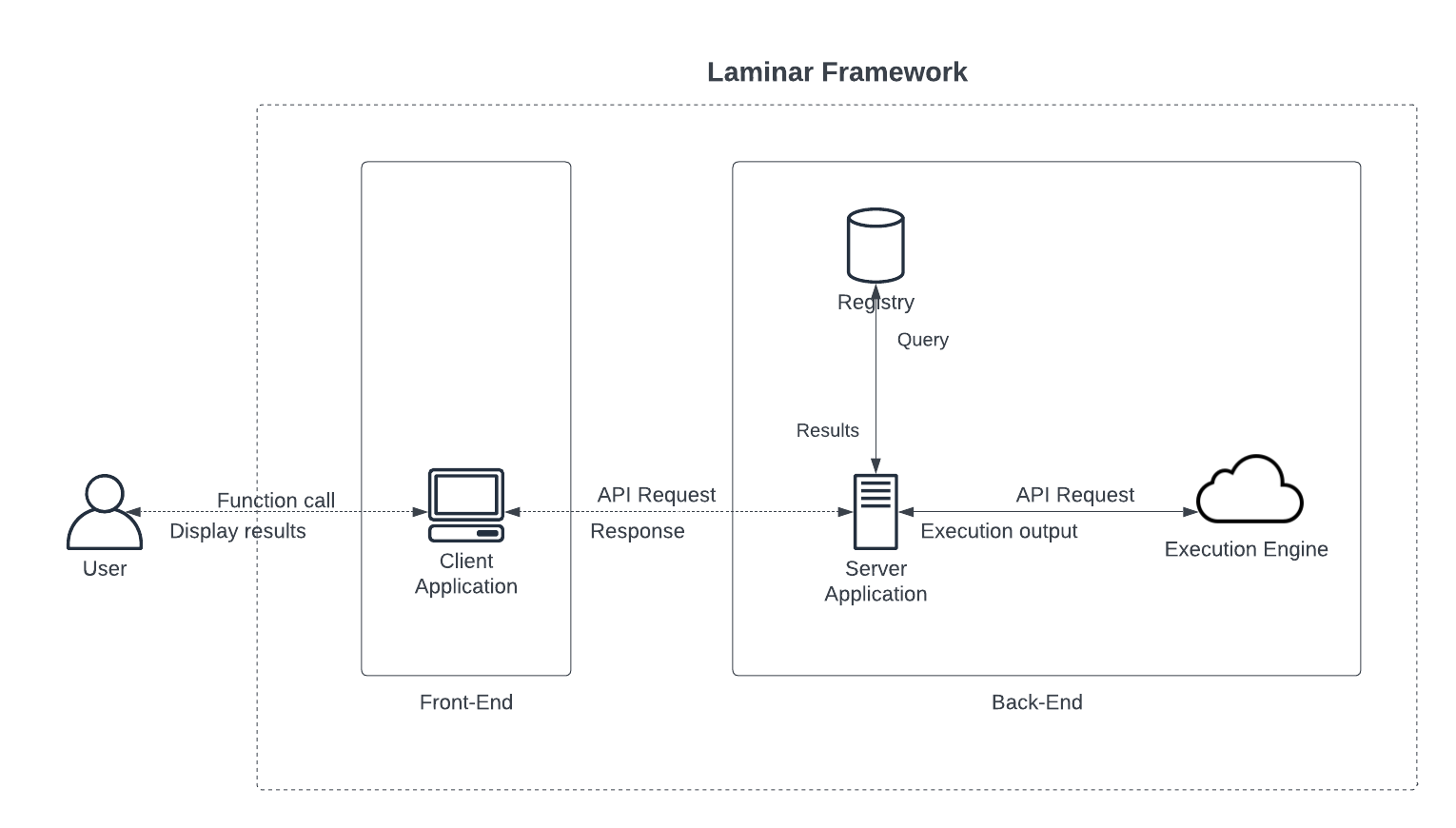}
\caption{Architectural Overview of \texttt{Laminar} }
\label{fig:system_arch}
\end{figure}

Depicted in Figure~\ref{fig:system_arch}, \texttt{Laminar} architecture distinctly separates the client, server, and execution functionalities. In this serverless architecture, the back-end responds to client-triggered events (captured by the front-end) by executing code within an isolated environment, like a remote computing infrastructure (e.g. Cloud system or HPC cluster). This design ensures scalability and ephemerality, with the environment being dismantled upon completion to optimize resource efficiency.

\vspace{-0.5cm}
%While conforming to the conventional Client-Server architecture, \texttt{Laminar} introduces an execution engine dedicated to handling serverless execution, establishing a streamlined pattern for communication. 

%The Client-Server model features clients sending requests to the server and receiving responses, while the execution engine exclusively interfaces with the server. The server provides services to one or more client applications, which make requests and receive responses from the server. The server is a
%centralized system that manages data and resources, while the clients consume the services provided by the server. By design, the \texttt{Laminar} framework leverages a Python Flask server as the execution engine, introduced later in Section ~\ref{section:execution}, which is the component responsible for executing streaming workflows in a serverless manner. 

. %\texttt{Laminar} uses a Python
%Flask server as an execution engine and runs Workflows in this environment, this is discussed further in Section ~\ref{section:execution}.

\subsection{Registry}
\label{section:registry}

\begin{figure}[h!]
  \centering
\includegraphics[width=\linewidth]{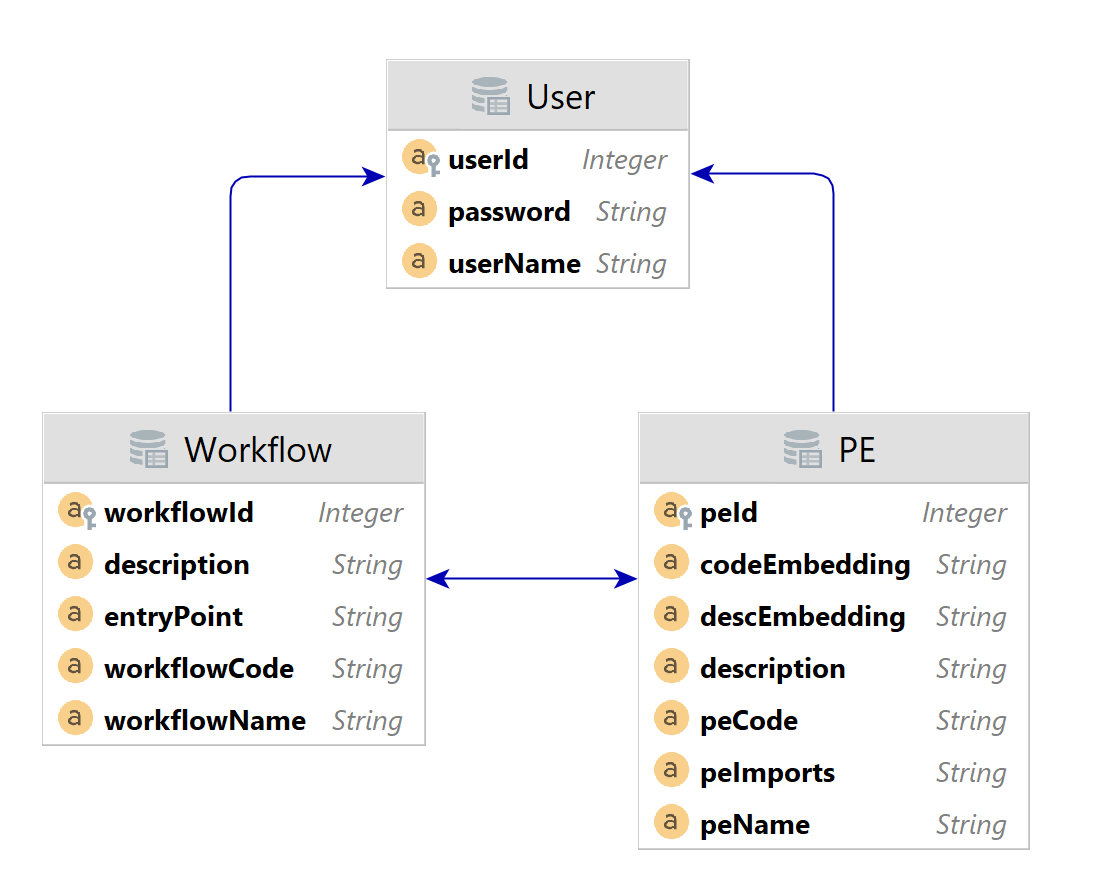}
\vspace{-0.4cm}
  \caption{Registry Schema}
  \label{fig:registry}
\end{figure}

The \textit{Registry}, hosted remotely on the web-based service~\footnote{\url{https://www.freesqldatabase.com/}}, serves as a central repository housing details about users, Processing Elements (PEs), and dispel4py workflows. It includes descriptions and various properties for each entity, as depicted in Figure~\ref{fig:registry} and Table~\ref{table:entity-properties}. Users are associated with both PEs and workflows through a one-way many-to-many relationship, ensuring that they are linked to their respective registered PEs or workflows while maintaining data privacy and preventing unauthorized access to information. This design approach promotes the concept of PE and workflow "owners" within the unified registry system, eliminating the need for individual registries. When a user intends to register a PE that is already associated with another user, the system includes the user as an additional user (or owner) of the PE or workflow, rather than creating a duplicate entry. This process ensures that users can efficiently access their registered PEs or workflows without redundant data entries.

\vspace{-0.4cm}

\begin{table}[h!]
\centering
\small
\begin{tabularx}{\linewidth}{|>{\raggedright\arraybackslash}p{1cm}|X|}
\hline
\textbf{Entity} & \textbf{Properties} \\
\hline
PE & 
\textbf{peId}: A unique identifier for a PE entry \\
& \textbf{codeEmbedding}: The code embedding of a PE \\
& \textbf{descEmbedding}: The description embedding of a PE \\
& \textbf{description}: A field indicating the PE's functionality, provided by the user or automatically summarized \\
& \textbf{peCode}: The serialized code for a PE \\
& \textbf{peImports}: Import dependencies for a PE \\
& \textbf{peName}: The class name of a PE \\

\hline

Workflow & 
\textbf{workflowId}: A unique identifier for a Workflow entry \\
& \textbf{description}: An optional field describing the functionality of a Workflow \\
& \textbf{entryPoint}: A unique name identifier for a Workflow \\
& \textbf{workflowCode}: The serialized code for a Workflow \\
& \textbf{workflowName}: The class name for a Workflow \\
\hline

User & 
\textbf{userId}: A unique identifier for a user entry \\
& \textbf{password}: User password \\
& \textbf{userName}: A unique name identifier for a user \\
\hline
\end{tabularx}
\caption{Registry Entities and Properties}
\label{table:entity-properties}
\vspace{-0.4cm}
\end{table}

%\vspace{-0.4cm}

The relationship between PEs and workflows is established as a two-way many-to-many association. This design allows a PE to be associated with multiple workflows, and conversely, a workflow can consist of numerous PEs. This structure proves beneficial in terms of data duplication and enhanced querying capabilities. For instance, users often require the ability to identify all PEs belonging to a specific workflow. With the current setup, retrieving all PEs associated with a workflow becomes straightforward, simplifying the querying process. This efficient relationship design streamlines user interactions with the system's API, as data pertaining to a workflow can be accessed without the need for additional work.

%In terms of physical implementation, the Java Persistence API (JPA)
%was used to create the registry. When the Spring Boot Server application is started (See Section ~\ref{section:server}), the JPA provider (Hibernate) will automatically generate the necessary SQL statements to build the database. The JPA EntityManager API is then used to perform CRUD operations on the database.

\subsubsection{PE Summarization and Embeddings}
\label{section:regitry-summarization}

To enable seamless semantic code search and code completion, the \textit{Registry} incorporates PE summarizations. These summarizations are generated by the \textit{Client} (refer to Section~\ref{section:PEs-semanticSearch}) when users do not provide a description while registering a PE. As introduced in Table~\ref{table:entity-properties}, the PE \textbf{description} property can include either the user-provided description or an automatically generated summary using the \textit{CodeT5} Language model. These PE summaries and descriptions form the cornerstone of our semantic code search functionality.

Additionally, the \textit{Registry} includes storage for PE code embeddings (\textbf{codeEmbedding} property) and PE description embeddings (\textbf{descEmbedding} property). These embeddings are generated by the \textit{Client} (as discussed in Sections~\ref{section:code-completion} and~\ref{section:PEs-semanticSearch}) during PE registration, contributing to enhanced performance outcomes. Storing these embeddings allows us to perform efficient semantic code searches and code completions without the need to re-calculate them every time a user initiates a search. This re-use of embeddings significantly enhances the responsiveness and performance of our system, ensuring a seamless user experience while navigating and querying the \textit{Registry}.

\subsection{Server}
\label{section:server}

The \textit{Server} is a foundational element within the \texttt{Laminar} architecture, housing the core domain logic. Organized using the layered design pattern, it consists of distinct tiers: \textit{Controller}, \textit{Service}, \textit{Model}, and \textit{Data Access Object} (\textit{DAO}) layers. These layers, pivotal to the architecture, collectively empower the \textit{Server} with modular and structured functionality.

\subsubsection{Controller Layer}

At the top, the \textit{Controller layer} acts as an interface, handling client requests and orchestrating responses using JSON for data exchange. The \textit{Laminar API} in this layer serves as the conduit for client-server interaction. We have different controllers for different parts of the system (e.g. PE, Workflow, Execution, etc.) Table~\ref{table:api-controllers} enumerates the API endpoints for each controller. 

\vspace{-0.1cm}

\begin{table}[h!]
\centering
\footnotesize
\begin{tabularx}{\linewidth}{|>{\raggedright\arraybackslash}p{1.2cm}|X|}
\hline
\textbf{Controller} & \textbf{Endpoints} \\
\hline
PE & \texttt{POST /registry/\{{user\}}/pe/add} \\ 
& \texttt{GET /registry/\{{user\}}/pe/all} \\
& \texttt{GET /registry/\{{user\}}/pe/id/\{{id\}}} \\
& \texttt{GET /registry/\{{user\}}/pe/name/\{{name\}}} \\
& \texttt{DELETE /registry/\{{user\}}/pe/remove/id/\{{id\}}} \\
& \texttt{DELETE /registry/\{{user\}}/pe/remove/name/\{{name\}}} \\
\hline

Workflow & \texttt{POST /registry/\{{user\}}/workflow/add} \\ 
& \texttt{GET /registry/\{{user\}}/workflow/all} \\
& \texttt{GET /registry/\{{user\}}/workflow/id/\{{id\}}} \\
& \texttt{GET /registry/\{{user\}}/workflow/name/\{{name\}}} \\
& \texttt{GET /registry/\{{user\}}/workflow/pes/id/\{{id\}}} \\
& \texttt{GET /registry/\{{user\}}/workflow/pes/name/\{{name\}}} \\
& \texttt{DELETE /registry/\{{user\}}/workflow/remove/id/\{{id\}}} \\
& \texttt{DELETE /registry/\{{user\}}/workflow/remove/name/\{{name\}}} \\
& \texttt{PUT /registry/\{{user\}}/workflow/\{{workflowId\}}/pe/\{{peId\}}} \\
\hline

Execution & \texttt{POST /execution/\{{user\}}/run} \\
\hline

Registry & \texttt{GET /registry/\{{user\}}/all} \\
& \texttt{GET /registry/\{{user\}}/search/\{{search\}}/type/\{{type\}}} \\
\hline

User  & \texttt{GET /auth/all} \\
& \texttt{POST /auth/login} \\
& \texttt{POST /auth/register} \\
\hline
\end{tabularx}
\caption{\textit{Laminar API} Controllers and Endpoints} 
\label{table:api-controllers}
\vspace{-0.4cm}
\end{table}

\vspace{-0.4cm}

\subsubsection{Service Layer}

Beneath, the \textit{Service layer} houses the application's business logic. Here, data from the controller is processed, detached from specific data storage details, and reliant on the DAO layer for data access.

\subsubsection{Data Access Object (DAO) layer}
The \textit{DAO layer} interacts with the data store, executing essential CRUD operations—create, read, update, delete—ensuring seamless data management. 

\subsubsection{Model Layer}
Lastly, the \textit{Model layer} introduce an object-oriented representation of system data, including \textit{Registry} entity definitions introduced in Section~\ref{section:registry}. 

\subsubsection{Error Handling}

To further augment server-side operations, tailored error handling has been implemented in the \textit{Server} across layers to address unforeseen and unauthorized behaviors, encompassing scenarios such as invalid login credentials. These meticulously designed exceptions furnish critical insights, including type identification, error codes, failed parameters, and supplementary details, thereby enhancing user comprehension and overall experience. Additionally, conformity to a standardized JSON format streamlines exception response formatting on the client side.

\subsection{Execution Engine}
\label{section:execution}

The \textit{Execution Engine} is the serverless core of \texttt{Laminar}, enabling remote execution of workflows ( or just singel PEs) with a single API endpoint: \texttt{/execution/{user}/run} (see Table~\ref{table:api-controllers}). This endpoint is the gateway for execution requests, including workflows, PEs, runtime configs, arguments, imports, and mappings. 

Within a conda Python environment~\footnote{A conda environment is a self-contained directory that holds specific software packages and their dependencies, allowing for easy management and isolation of software environments.}, the execution engine is furnished with the \texttt{dispel4py} library and its essential packages. This guarantees a smooth execution environment, sparing users from remote environment management. An intelligent auto-import mechanism scrutinizes the varied import dependencies of streaming workflows, creating an all-inclusive requirement list transmitted to the \textit{Execution Engine}. It autonomously imports necessary prerequisites, eliminating the need for user installations. 

 The \textit{Execution Engine} handles additional resources essential for workflow execution, such as data from various sources. Users can compile these resources in a 'resources' directory, managed by the application for smooth serialization and deserialization during execution. An additional enhancement is the autonomous identification of the initial PE within a workflow, crucial for invoking the mapping function that initiates workflow execution. While conventional \texttt{dispel4py} often requires users to specify the initial PE, the \textit{Execution Engine} autonomously analyzes the workflow's structure to identify the suitable starting point, minimizing user involvement in complex tasks, aligned with the \textit{Client}'s design philosophy

Leveraging \texttt{dispel4py}, \texttt{Laminar} harnesses enhanced capabilities. The \textit{Execution Engine} enables streamlined parallel execution of streaming workflows through diverse \texttt{dispel4py} parallel mappings. Users effortlessly select mappings via the \textit{Client}, enhancing application speed and resource utilization.

In the future we plan to expand \texttt{Laminar}'s capabilities by enabling the registration of multiple \texttt{Execution Engines}, a process that currently involves manual intervention. This  will significantly improve convenience and management efficiency for users.

\subsection{Client}
\label{section:client}

\begin{figure}[h!]
  \centering
\includegraphics[width=\linewidth]{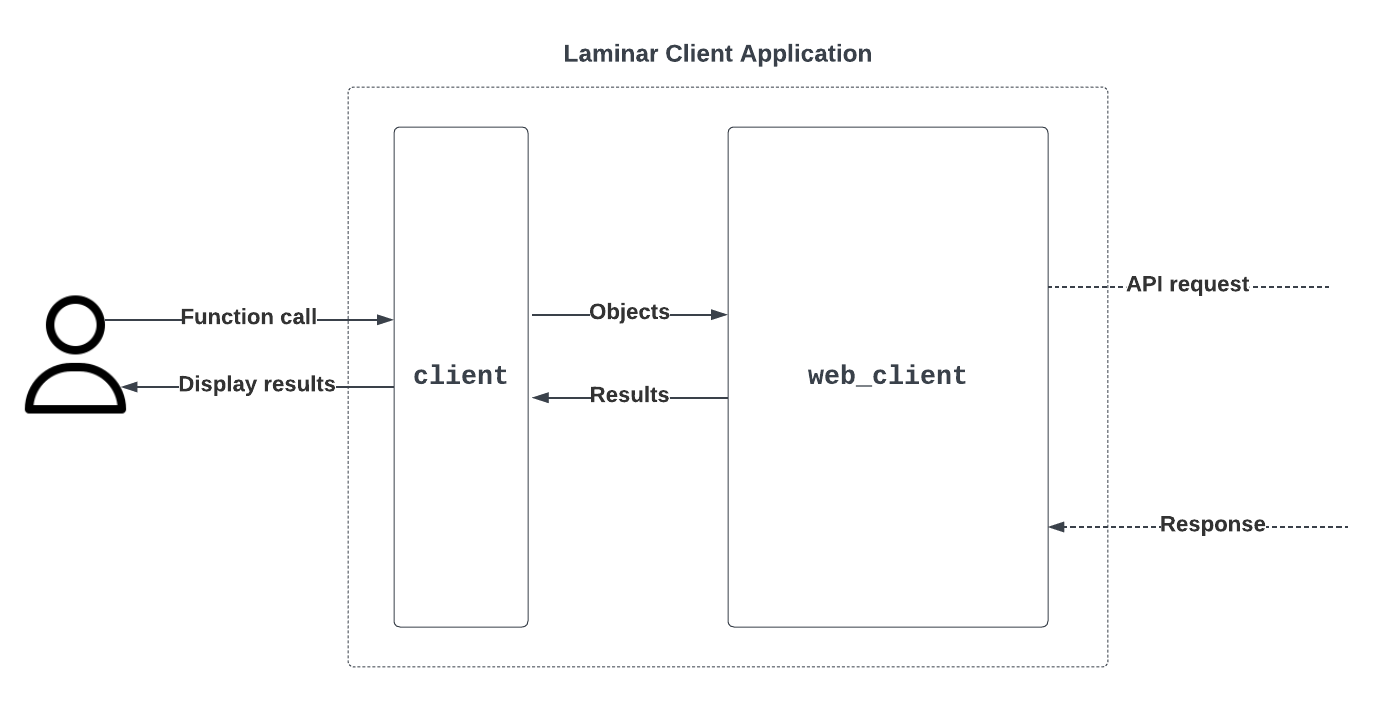}
 %\vspace{-0.4cm}
  \caption{Dual-Layer Client Structure in \texttt{Laminar} Framework}
  \label{fig:client-architecture}

\end{figure}

The \textit{Client} component serves as a user-friendly Python application, designed to ensure smooth engagement with the \texttt{Laminar} framework. It adopts a dual-layer structure as is illustrated in Figure~\ref{fig:client-architecture} comprising the \textit{client layer} and the \texttt{web\_client} layer, enhancing accessibility and ease of use. As follows both layers are explained.

%\vspace{-0.4cm}

\subsubsection{Client Layer}
\label{section:client-layer}
This layer offers user-accessible functions for tasks like registering PEs or workflows or executing workflows in a serverless manner. Currently, we support the following functions~\footnote{Full user manual with more examples can be found here \url{https://tinyurl.com/355zps8p}}:

\vspace{0.2cm}

\begin{enumerate}
\small
\item \textbf{register:} User cans register with a name and a password.

\begin{minted}[mathescape,
               numbersep=5pt,
               autogobble,
               frame=single,
               framesep=1mm,
               breaklines,
               breakafter=_]{python}
   register(self, user_name:str, user_password:str)
\end{minted}

Example:

\begin{minted}[mathescape,
               numbersep=5pt,
           autogobble,
           frame=single,
           framesep=1mm]{python}
    client.register("zz46","password")
\end{minted}

\item \textbf{login:} Users can  login with their details.

\begin{minted}[mathescape,
           numbersep=5pt,
           autogobble,
           frame=single,
           framesep=1mm]{python}
login(self,user_name:str,user_password:str)
\end{minted}
Example:
\begin{minted}[mathescape,
           numbersep=5pt,
           autogobble,
           frame=single,
           framesep=1mm]{python}
    client.login("zz46","password")
\end{minted}

\item \textbf{register\_PE:} Users can register PEs to store in the \textit{Registry}.

\begin{minted}[mathescape,
           numbersep=5pt,
           autogobble,
           frame=single,
           framesep=1mm]{python}
register_PE(self,pe:PE_TYPES,description:str=None)        
\end{minted}

Example:
\begin{minted}[mathescape,
           numbersep=5pt,
           autogobble,
           frame=single,
           framesep=1mm]{python}
    client.register_PE(NumberProducer,
         "Random numbers producer")
\end{minted}

\item \textbf{register\_Workflow:} Similarly, workflows can also be registered.

\begin{minted}[mathescape,
           numbersep=5pt,
           autogobble,
           frame=single,
           framesep=1mm]{python}
register_Workflow(
    self, workflow: WorkflowGraph,
    workflow_name:str,
    description:str=None
)
\end{minted}

Example:
\begin{minted}[mathescape,
           numbersep=5pt,
           autogobble,
           frame=single,
           framesep=1mm]{python}
    client.register_Workflow(
        graph, 
        "isPrime",
        "Workflow that prints random prime numbers"
    )
\end{minted}

\item \textbf{remove\_PE:} PEs in the \textit{Registry} can be deleted if no longer required.

\begin{minted}[mathescape,
           numbersep=5pt,
           autogobble,
           frame=single,
           framesep=1mm]{python}
remove_PE(self,pe:Union[str,int])
\end{minted}

Example:
\begin{minted}[mathescape,
           numbersep=5pt,
           autogobble,
           frame=single,
           framesep=1mm]{python}
    client.remove_PE("NumberProducer")
\end{minted}

\item \textbf{remove\_Workflow:} Users can delete workflows from the \textit{Registry}.

\begin{minted}[mathescape,
           numbersep=5pt,
           autogobble,
           frame=single,
           framesep=1mm]{python}
remove_Workflow(self,workflow:Union[str,int])
\end{minted}

Example:
\begin{minted}[mathescape,
           numbersep=5pt,
           autogobble,
           frame=single,
           framesep=1mm]{python}
    client.remove_Workflow("IsPrime")
\end{minted}

\item \textbf{get\_PE:} registered PEs can be retrieved for creating new workflows.

\begin{minted}[mathescape,
           numbersep=5pt,
           autogobble,
           frame=single,
           framesep=1mm]{python}
get_PE(self,pe:Union[str,int],describe:bool=False)
\end{minted}

Example:
\begin{minted}[mathescape,
           numbersep=5pt,
           autogobble,
           frame=single,
           framesep=1mm]{python}
        pe1 = client.get_PE("NumberProducer")
\end{minted}

\item \textbf{get\_Workflow:} workflows can be retrieved for execution.

\begin{minted}[mathescape,
           numbersep=5pt,
           autogobble,
           frame=single,
           framesep=1mm]{python}
get_Workflow(
    self,
    workflow:Union[str,int],
    describe:bool = False
)
\end{minted}
Example:
\begin{minted}[mathescape,
           numbersep=5pt,
           autogobble,
           frame=single,
           framesep=1mm]{python}
     graph = client.get_Workflow("IsPrime")
\end{minted}
\item \textbf{get\_PEs\_By\_Workflow:} Users can get a list of PEs for a  workflow.

\begin{minted}[mathescape,
           numbersep=5pt,
           autogobble,
           frame=single,
           framesep=1mm]{python}
get_PEs_By_Workflow(self,workflow:Union[str,int])
\end{minted}

Example:
\begin{minted}[mathescape,
           numbersep=5pt,
           autogobble,
           frame=single,
           framesep=1mm]{python}
    pes = client.get_PEs_By_Workflow("IsPrime")
\end{minted}

\item \textbf{search\_Registry:} Search \textit{Registry} for PEs and workflows.  \textbf{Section~\ref{section:searches}}. 

\begin{minted}[mathescape,
           numbersep=5pt,
           autogobble,
           frame=single,
           framesep=1mm]{python}
search_Registry(
self, 
search:str,
search_type:_TYPES = "both", 
query_type:_QUERY_TYPES = "text")
\end{minted}

Example: 
\begin{minted}[mathescape,
           numbersep=5pt,
           autogobble,
           frame=single,
           framesep=1mm]{python}
    results= client.search_Registry("isPrime", "workflow")
\end{minted}

\item \textbf{describe:} It provides information on PEs and workflows based on name and description properties.

\begin{minted}[mathescape,
           numbersep=5pt,
           autogobble,
           frame=single,
           framesep=1mm]{python}
describe(self, obj:any)
\end{minted}

Example: 
\begin{minted}[mathescape,
           numbersep=5pt,
           autogobble,
           frame=single,
           framesep=1mm]{python}
    client.describe(IsPrime)
\end{minted}

\item \textbf{get\_Registry:} Retrieves a list of all items stored in the \textit{Registry}.

\begin{minted}[mathescape,
           numbersep=5pt,
           autogobble,
           frame=single,
           framesep=1mm]{python}
get_Registry(self)
\end{minted}

Example:
\begin{minted}[mathescape,
           numbersep=5pt,
           autogobble,
           frame=single,
           framesep=1mm]{python}
    registry = client.get_Registry()
\end{minted}

\item \textbf{run:} Users can execute workflows at the \textit{Execution Engine}.

\begin{minted}[mathescape,
           numbersep=5pt,
           autogobble,
           frame=single,
           framesep=1mm]{python}
#process parameter indicates the mapping
# It accepts: SIMPLE, MULTI, MPI, REDIS. 

#input indicates the number of iterations
#for which the workflow will be running
run(
    self, workflow:Union[str,int,WorkflowGraph],
    input=None,
    process=:_MAPPING_TYPES = "SIMPLE",
    args=None,
    resources:bool=False)
\end{minted}

 Example:
\begin{minted}[mathescape,
           numbersep=5pt,
           autogobble,
           frame=single,
           framesep=1mm]{python}
     #Simple mapping is automatically inferred
     # when no mapping is specified.
     # Running the workflow for 5 iterations
     client.run("IsPrime", input=5)
\end{minted}

\end{enumerate}

Note that users possess the flexibility to submit workflows using the \texttt{run} function from the client (as demonstrated above), which can involve multiple PEs as showcased in the use cases outlined in Section~\ref{section:use-cases}. Alternatively, they have the option to create workflows with a single PE, such as a \textit{ProducerPE} or \textit{GenericPE}, similar to traditional FaaS frameworks. PEs, whether as components of a workflow or individual units run through \texttt{Laminar}, can demonstrate either stateful behavior like the \texttt{CountWords} PE in Listing~\ref{code:countPE}, or stateless behavior, as exhibited by the \texttt{NumberProducer} in Listing~\ref{code:numberproducer}.

\vspace{-0.25 cm}

\begin{listing}[h]
\begin{minted}[mathescape,
               numbersep=5pt,
               gobble=2,
               frame=single,
               framesep=1mm]{python}
  class CountWords(GenericPE):
      def __init__(self):
          from collections import defaultdict
          GenericPE.__init__(self)
          #Add an input port named "input", from which
          #it will receive tuples with shape (word, 1).
          #Data is group-by (MapReduce) 
          #the first element (index 0) of the tuples 
          self._add_input("input", grouping=[0])
          
          #Add an output port named "output"
          self._add_output("output")
          
          #Initialize a stateful variable 
          #to store word counts
          self.count = defaultdict(int)
        
      def _process(self, inputs):
          import os 
          #Extract word and count from the input
          word, count = inputs['input']
          # Update the count for the word
          self.count[word] += count
\end{minted}
\vspace{-0.2cm}
\caption{\texttt{Stateful PE using group-by for word count}}
\label{code:countPE}
\end{listing}

\subsubsection{Web Client Layer}
\label{section:web_client-layer}

This layer holds an important role within the \texttt{Laminar} framework. This intermediary layer acts as a conduit for communication, orchestrating the flow of data and interactions between the client-side processing and the server-side execution, as is shown in Figure~\ref{fig:client-architecture}. A significant aspect of this layer is its versatility in handling multiple input types. For instance, the \texttt{workflow} parameter of the \texttt{run} function introduced in ~\ref{section:client-layer} can accept various input formats such as a \texttt{workflow name} (\texttt{str}),  a \texttt{workflow ID} (\texttt{int}), or \texttt{workflow object} {\texttt{WorkflowGraph}} for execution  - \texttt{workflow:Union[str,int,WorkflowGraph]}. Combining these capabilities enhances user experience and simplifies usage. 

A notable challenge that we took in this layer was the code serialization – a crucial step to package Workflows and PEs in a format comprehensible to the execution engine. To tackle this, external packages were utilized for serialization. \textit{cloudpickle} library \footnote{\url{https://github.com/cloudpipe/cloudpickle}}, chosen after evaluating alternatives like \textit{pickle}\footnote{\url{https://docs.python.org/3/library/pickle.html}} and \textit{dill}\footnote{\url{https://pypi.org/project/dill/}}, emerged as the preferred option. Its capability to serialize complex Python objects, including classes and recursive structures, proved essential. Furthermore, cloudpickle's suitability for transmitting code over networks to remote hosts aligned with the project's needs.

Serialization also demanded consideration for storage in the \textit{Registry}. The serialized code, presented as a byte string, needed a suitable format for storage. To ensure portability, a \texttt{base64} encoding\footnote{\url{https://docs.python.org/3/library/base64.html}} was applied to convert the byte stream into a string format. This serialization approach was leveraged consistently across various code segments required for execution.

This layer also addresses the challenge of handling dependencies. While \textit{cloudpickle} handles import dependencies, an extra library, \texttt{findimports}\footnote{\url{https://pypi.org/project/findimports/}}, was used to analyze classes for imports. Users are required to specify necessary imports within PEs to ensure the execution environment has all the required imports for a successful workflow execution.  In summary, the \textit{web\_client} layer forms the core for  client-server interactions and smooths computations throughout \texttt{Laminar}.

\section{Registry Search and Exploration}
\label{section:searches}

The \textit{Client} offers a comprehensive set of search and exploration functionalities within the registry, facilitating efficient retrieval and discovery of stored PEs and workflows. This section outlines the different search mechanisms available within \texttt{Laminar} framework.

\subsection{Text-Based Search}

Text-based searches empower users to quickly locate relevant workflows based on textual information. The \textit{Client} adeptly processes user-input text queries and effectively matches them with workflows based on their names and descriptions. This functionality includes support for partial matching\footnote{User input text and stored workflow textual information are normalized during a preprocessing step.}, enabling instances where a user (see Figure~\ref{fig:search-text-based}) queries `prime' and the system successfully identifies a registered workflow named `isPrime' (workflow ID 2), thereby enhancing workflow search accuracy and user convenience.

\vspace{0.2cm}
\begin{minted}[mathescape,
           numbersep=5pt,
           autogobble,
           frame=single,
           framesep=1mm]{python}
      client.search_Registry("prime", "workflow")
\end{minted}
\par Output:
\vspace{-0.2cm}
\begin{figure}[h!]
\includegraphics[scale=0.3]{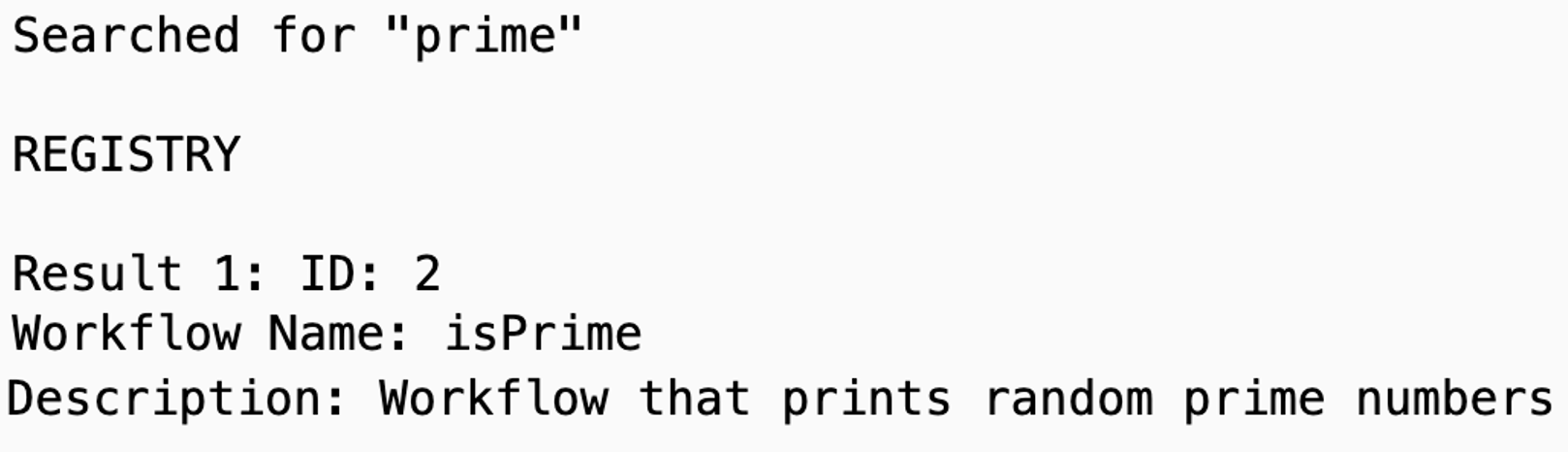}
\vspace{-0.4cm}
  \caption{Text-based search for workflows containing the term 'prime' in their names or descriptions. The result is the `isPrime' workflow with ID `2'.}
  \label{fig:search-text-based}
\end{figure}

\vspace{-0.5cm}
\subsection{Semantic Code Search}
\label{section:PEs-semanticSearch}

In the domain of semantic code search, \texttt{Laminar}\textit{Client} employs a sophisticated process to enable contextually-aware searches for Processing Elements (PEs) based on their user-provided descriptions. This powerful capability enables the discovery of PEs that align with user requirements. If no description is provided during PE registration, the \texttt{Client} employs a workaround: it employs code summarization to automatically generate  summaries based on the PE's code itself. This automatic summarization utilizes the \multisum{} model\footnote{A fine-tuned variant of the \texttt{Code T5} model, trained on the CodeSearchNet dataset, available at \url{https://huggingface.co/Salesforce/codet5-base-multi-sum}}, which has been evaluated in comparison to other models for Python code summarization in a recent study~\cite{repograph}, resulting in the selection of this model due to its superior performance. 

Whether manually input or automatically generated, PE descriptions are transformed into high-dimensional vectors that encapsulate their semantic information, facilitating efficient similarity analysis. This is accomplished using the fine-tuned \textit{unixcoder-code-search} model\footnote{\url{https://huggingface.co/Lazyhope/unixcoder-nine-advtest}}, specialized for code search tasks, as detailed in~\cite{unixcoder}. Our team fine-tuned this model as an extension of our work ~\cite{reposim}, adhering to the guidelines outlined 
in~\footnote{\url{https://github.com/microsoft/CodeBERT/tree/master/UniXcoder/downstream-tasks/}}, 
and leveraging the AdvTest dataset ~\cite{codexglue}. Its performance was benchmarked against the Unixcoder base model (see Section~\ref{section:ev-semantic-search}).

It is noteworthy that the embeddings for PE descriptions are calculated just once, during the registration process. These embeddings are then stored within the \textit{Registry} (refer to Section~\ref{section:registry}) under the \textbf{descEmbedding} property.

Upon user query submission, \texttt{Laminar} transforms the query using the \textit{unixcoder-code-search} model, calculating cosine similarity against all registered PE description embeddings. This approach, rooted in the bi-encoders paradigm (explained in Section~\ref{section:code-search}), enhances search accuracy and context. The depicted scenario in Figure~\ref{fig:semantic-code-search} showcases a user who has registered 22 PEs and five workflows (two detailed in Section~\ref{section:use-cases}). In this instance, a semantic search for the text `A PE that checks if a number is prime` is conducted, resulting in ranked PEs based on similarity scores.
\vspace{0.2cm}
\begin{minted}[mathescape,
           numbersep=5pt,
           autogobble,
           frame=single,
           framesep=1mm,
           escapeinside=||]{python}
client.search_Registry("A PE that checks if 
                       |\textcolor{FireBrick}{a number is prime}|",
                       "pe","text")
\end{minted}

\par Output: 

\begin{figure}[h!]
\includegraphics[scale=0.25]{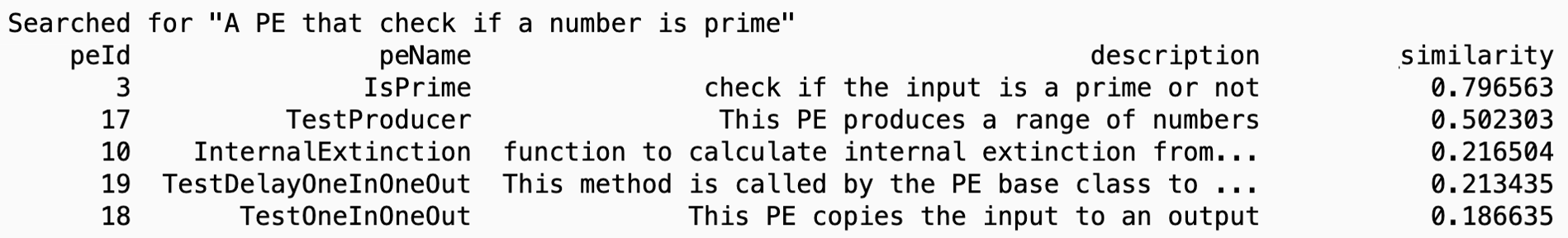}
\caption{Semantic search of registered PEs. Automated descriptions generated by \texttt{Laminar} for PEs with peIDs 3 and 10}
  \label{fig:semantic-code-search}
\end{figure}

This feature enhances the user experience, allowing natural language input to be intelligently matched with relevant PEs. Users later can retrieve (e.g using \texttt{get\_PE()} client function) a registered PE for creating new workflows. 

%As a result, users are presented with a carefully curated list of potential PE candidates, enhancing the efficiency and confidence with which workflows are constructed.

\subsection{Code Completion}
\label{section:code-completion}

Efficient and accurate code completion forms a pivotal cornerstone in the toolkit of not only developers but also scientists and research software engineers. \texttt{Laminar} integrates the \textit{ReACC-py-retriever} model (introduced in Section~\ref{section:LM}) to tackle code completion tasks head-on. This functionality empowers users to explore and retrieve relevant Processing Elements (PEs) based on their input. This input can be a partial or complete code query for a specific PE. Our selection of the \textit{ReACC-py-retriever} model is the result of an evaluation against alternative models for the same task (refer to Section~\ref{section:ev-code-completion} for details).

Much like the process for semantic code search, the \textit{Client} constructs embeddings for each PE's code using the \textit{ReACC-py-retriever} model. These embeddings are subsequently stored in the registry, thoughtfully organized under the \textbf{codeEmbedding} property.

When a user seeks to execute a query using a code snippet relevant to a specific PE, the \textit{Client} generates an embedding for the provided query code snippet, employing the same \textit{ReACC-py-retriever} model. The  \texttt{Laminar} calculates the cosine similarity between the query's embedding and the embeddings of all registered PEs' codes. This  comparison leads to the identification of PEs whose code snippets closely resonate with the user's query.

\texttt{Laminar} enhances the coding experience by offering relevant and contextually appropriate PEs. The system accommodates code snippets that may not necessarily warrant completion; they might manifest as fragments of functionality. \texttt{Laminar} extracts the most relevant or fitting code snippet to augment the user's input. 

 \begin{minted}[mathescape,
           numbersep=5pt,
           autogobble,
           frame=single,
           framesep=1mm]{python}
client.search_Registry("random.randint(1, 1000)",
                        "pe","code")
\end{minted}
\par Output: 
\begin{figure}[h!]
\includegraphics[scale=0.25]{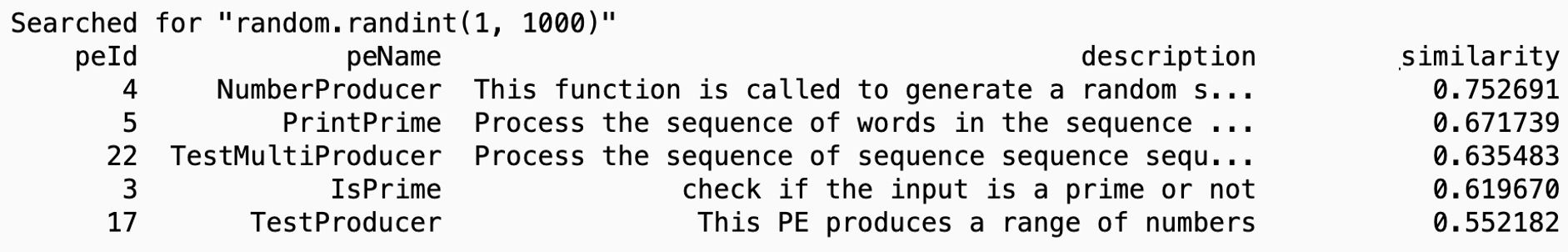}
  \caption{Semantic code completion for registered PEs relevant to the code "random.randint(1, 1000)"}
  \label{fig:code-completion-search}
\end{figure}

Figure~\ref{fig:code-completion-search} shows the process of semantic code completion for registered Processing Elements (PEs) is illustrated. The scenario involves searching for PEs that are relevant to the code snippet \texttt{random.randint(1, 1000)}. This code snippet serves as a query for code completion, seeking registered PEs whose code segments are similar to the provided snippet. The results are presented in ascending order of similarity score.

Subsequently, users can retrieve their desired PEs, such as the one with the highest similarity score, to incorporate it into a new workflow or create a new PE by reusing code segments. This feature expedites development, eases cognitive load, and notably minimizes the chances of errors during new PE creation.

\section{Computational Showcases}
\label{section:use-cases}

In this section we are going to introduce two \texttt{dispel4py} workflows to showcase \texttt{Laminar} functionality. 

\subsection{IsPrime workflow}
\label{section:IsPrime}

The \texttt{IsPrime} workflow (Listing\ref{listing:workflow-code}) continuously streams random numbers using the `NumberProducer' PE. It assesses their primality with the `IsPrime' PE and prints the prime numbers using `PrintPrime' PE. This exemplifies data flow and processing, connecting PEs to identify and print prime numbers. The workflow is depicted by the green graph in Figure~\ref{fig:dispel4py}.

\begin{figure}[h!]
\includegraphics[scale=0.25]{
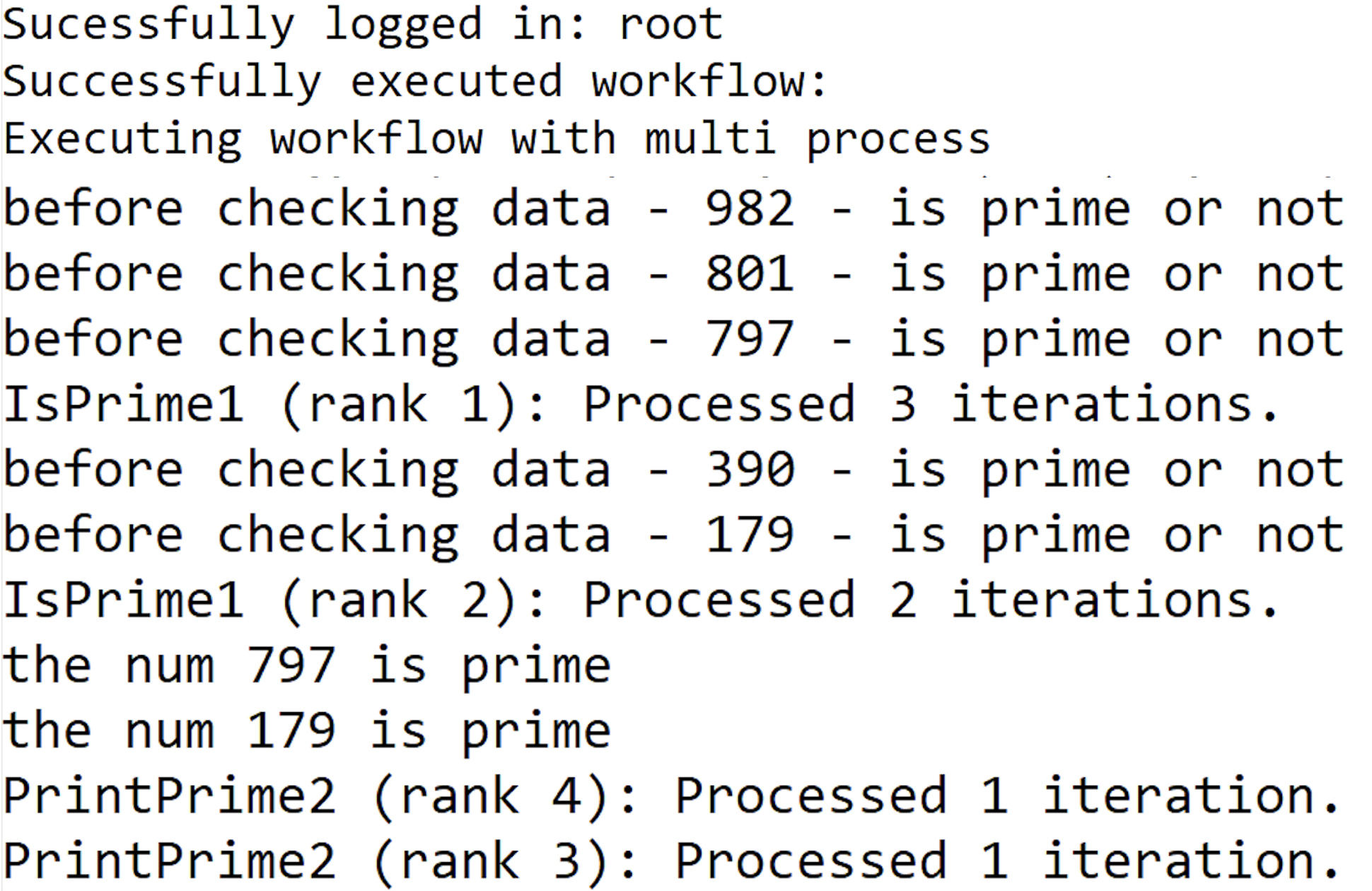}
\vspace{-0.5cm}
\caption{Output sent from the \textit{Execution Engine} to the \textit{Client}}
\label{fig:outputPrime}
\end{figure}

\vspace{-0.2cm}

\begin{listing}[h!]
\begin{minted}[mathescape,
               numbersep=5pt,
               gobble=2,
               frame=single,
               framesep=1mm]{python}
    class NumberProducer(ProducerPE):
        def __init__(self):
            ProducerPE.__init__(self)   
        def _process(self):
            # Generate a random number
            result= random.randint(1, 1000)
            # Return the number as the output
            return result
\end{minted}

\begin{minted}[mathescape,
           numbersep=5pt,
           autogobble,
           frame=single,
           framesep=1mm,
           escapeinside=||]{python}
    class IsPrime(IterativePE):
        def __init__(self):
            IterativePE.__init__(self)
        def _process(self, num):
            print("before checking data - %s - \\
                  |\textcolor{FireBrick}{is prime or not}|" %num)
            #Check if the given input (num) is prime
            if all(num % i != 0 for \\
              i in range(2, num)):
                 #Only if the input is prime,
                 #the value is return as an output 
                 return num
\end{minted}

\begin{minted}[mathescape,
               numbersep=5pt,
               gobble=2,
               frame=single,
               framesep=1mm]{python}
    class PrintPrime(ConsumerPE):
        def __init__(self):
            ConsumerPE.__init__(self)
        def _process(self, num):
            # Print the input (num)
            print("the num %s is prime" % num)
\end{minted}

\begin{minted}[mathescape,
               numbersep=5pt,
               gobble=2,
               frame=single,
               framesep=1mm]{python}
    # Create instances of the defined PEs
    pe1 = NumberProducer()
    pe2 = IsPrime()
    pe3 = PrintPrime()
    
    #Create a workflow graph 
    graph = WorkflowGraph()
    
    #Construct the graph, connecting the PEs
    #and their inputs and outputs 
    graph.connect(pe1, \\
                'output', pe2, 'input')
    graph.connect(pe2, \\
               'output', pe3, 'input')
\end{minted}
\vspace{-0.2cm}
\caption{\texttt{IsPrime} workflow, which corresponds to the abstract (green) graph shown in Figure~\ref{fig:dispel4py}.}
\label{listing:workflow-code}
\end{listing}

\begin{listing}[h!]
\begin{minted}[mathescape,
               numbersep=5pt,
               gobble=2,
               frame=single,
               framesep=1mm]{python}               
    client.run(graph,input=5,process=MULTI,
                 args={'num':5})
\end{minted}
\caption{Executing \texttt{IsPrime} within \texttt{Laminar} using the specified parameters, which correspond to the concrete (blue) graph shown in Figure~\ref{fig:dispel4py} }
\vspace{-0.2cm}
\label{listing:workflow-execution}
\end{listing}
%\vspace{-0.4cm}

When executing the \texttt{IsPrime} workflow in \texttt{Laminar}, a mapping strategy must be specified. In this case (shown in Listing~\ref{listing:workflow-execution}), we use the \textit{Multi} (\texttt{process}) parallel mapping. The configuration involves five iterations (\texttt{input}) with five processes (\texttt{num}). The first PE generates five random numbers, of which only two are prime in this scenario. Importantly, the \texttt{run()} function streamlines not only serverless workflow execution but also automates the registration of the workflow and its PEs.

Upon receiving the workflow and its configuration, the \textit{Execution Engine} initiates execution with the specified mapping (\textit{Multi} in this case). This configuration automatically adjusts the concrete workflow based on the number of processes chosen, eliminating the need for user intervention. The resulting output (shown in Figure~\ref{fig:outputPrime}) is then forwarded to the \textit{Client} by the \textit{Execution Engine}.

\subsection{Astrophysics workflow: Internal Extinction}
\label{section:Astro}
This workflow has been implemented to calculate the extinction within the galaxies, which is a significant property in astrophysics~\citep{filgueira2015dispel4py1}. This property reflects the dust extinction of the internal galaxies and is used for measuring the optical luminosity\footnote{\url{http://amiga.iaa.es/p/1-homepage.html}}. This workflow is reusable since it can be regarded as a prior step for other complex tasks which require this property.

\begin{figure}[htbp]
\centerline{\includegraphics[width=0.48\textwidth]{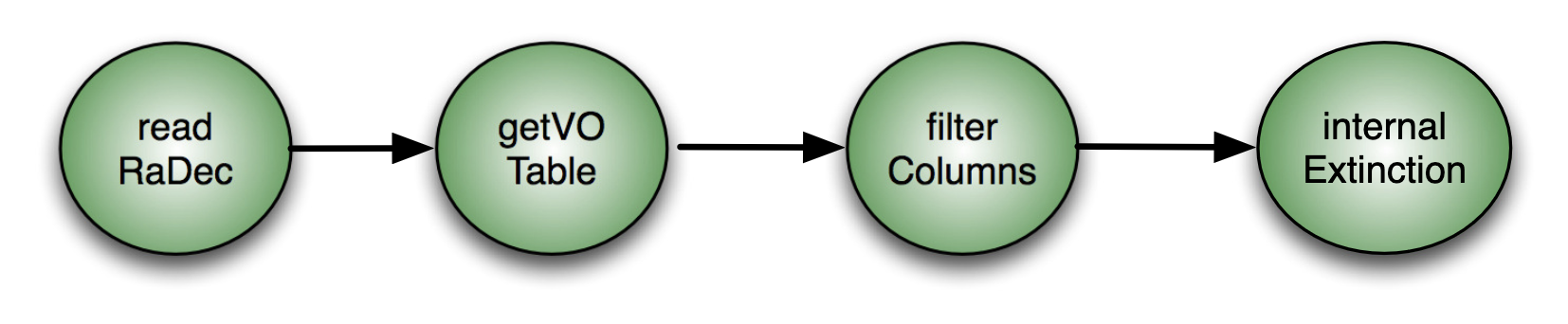}}
\vspace{-0.5cm}
\caption{Streaming workflow for calculating the internal extinction of galaxies.}
\label{fig:int}
\end{figure}

In Figure~\ref{fig:int}, the workflow involves four PEs. The \texttt{readRaDec} PE loads coordinate data from an input file, while \texttt{getVoTable} downloads a relevant VOTable from the Virtual Observatory website based on those coordinates. The \texttt{filterColumns} PE employs the astropy library\footnote{Python library for astronomy: \url{https://www.astropy.org/}} to parse and filter the VOTable data. Lastly, the \texttt{internalExt.} PE calculates internal extinction using data from the \texttt{Filt} PE. The complete workflow code is available here~\footnote{\url{https://github.com/dispel4pyserverless/dispel4py-client/blob/main/CLIENT\_EXAMPLES/AstroPhysics.py}}. Notably, \texttt{Laminar} detects and installs necessary imports for PEs, ensuring smooth operation within the \textit{Execution Engine}.
 
Listing~\ref{listing:workflow-registration}, provides an example of how the workflow has been stored with the \textit{Registry}. Following this registration, the workflow becomes accessible for retrieval at any desired moment, as depicted in Listing~\ref{listing:workflow-get}. Ultimately, the workflow can be executed within the serverless \textit{Execution Engine}, as demonstrated in Listing~\ref{listing:workflow-run}. For this instance, we have chosen to execute the workflow using the \textit{Redis} parallel mapping, utilizing ten processes for its execution.

\begin{listing}[h!]
\begin{minted}[mathescape,
           numbersep=5pt,
           autogobble,
           frame=single,
           framesep=1mm,
           escapeinside=||]{python}  
    client.register_Workflow(
        graph,
        "Astrophysics",
        "A workflow to compute the 
        |\textcolor{FireBrick}{internal extinction of galaxies}|")
\end{minted}

\vspace{-0.4cm}
\caption{Registering the \texttt{Astrophysics} workflow}
\label{listing:workflow-registration}
\end{listing}
\vspace{-0.2cm}
\begin{listing}[h!]
\begin{minted}[mathescape,
               numbersep=5pt,
               gobble=2,
               frame=single,
               framesep=1mm]{python}  
    workflow = client.get_Workflow("Astrophysics")
\end{minted}
\vspace{-0.3cm}
\caption{Retrieving the \texttt{Astrophysics} workflow}
\label{listing:workflow-get}
\end{listing}

%Note that the \textit{Redis} mappings also need the \texttt{redis\_ip} and \texttt{redis\_port} values to be specified. 
%\vspace{-0.2cm}

\begin{listing}[h!]
\begin{minted}[mathescape,
               numbersep=5pt,
               gobble=2,
               frame=single,
               framesep=1mm]{python}  
    client.run(workflow,
      input=[{"input":"resources/coordinates.txt"}],
      process=REDIS,
      args={'num':10}
      resources=True)
\end{minted}
\vspace{-0.2cm}
\caption{Executing the \texttt{Internal Extinction} workflow within \texttt{Laminar}, using \texttt{Redis} mapping and 10 processes.}
\label{listing:workflow-run}
\end{listing}

As mentioned in Section~\ref{section:execution}, \texttt{Laminar} supports additional resources that workflows may need. This is exemplified in the \texttt{Internal Extinction} workflow (Listing~\ref{listing:workflow-run}), where the required input file, containing relevant coordinate data (e.g., coordinates.txt'), is accessed from the resources' directory. Through a sequence of copying, serialization, and deserialization steps, the file becomes accessible to the \textit{Execution Engine} for use during workflow execution.

%Note that we can obtain all the PEs in a given registered workflow as shown Figure~\ref{figure:workflow-allPES}.

%\begin{minted}[mathescape,
%               numbersep=5pt,
%               gobble=2,
%               frame=single,
%               framesep=1mm]{python}  
%    list_of_pes = client.\\
%           get_PEs_By_Workflow("Astrophysics")
%\end{minted}
%\vspace{-0.2cm}

%\begin{figure}[H]
%\includegraphics[scale=0.3]{Figures/scenario3_output_1.png}
%\caption{Obtaining all the PEs of the \textit{Astrophysics} workflow.}
%\label{figure:workflow-allPES}
%\end{figure}

%\vspace{-0.2cm}

\section{Evaluation}
\label{section:evaluation}

\subsection{\texttt{Laminar} Performance}

To evaluate the performance of \texttt{Laminar}, a latency analysis was conducted, focusing on both the \textit{Simple} (sequential) and \textit{Multi} mappings (using 5 processes) of the \textit{Internal Extinction} workflow introduced in Section~\ref{section:Astro}. This evaluation aimed to compare the execution times between the original \texttt{dispel4py}  and \texttt{Laminar}, utilizing both local and remote \textit{Execution Engines}. Notably, \texttt{Laminar} accommodates both local and remote \textit{Execution Engines} with some manual adjustments, although in the future we will allow users to register multiple \textit{Execution Engines} from the \textit{Client}.

For the remote deployment scenario, we encapsulated the \textit{Execution Engine} within a Docker image~\footnote{\url{https://hub.docker.com/r/zz46/execution}}, tailored for easy deployment on diverse cloud platforms. In this experiment, we used the Azure Container Registry~\footnote{\url{https://azure.microsoft.com/en-gb/products/container-registry}} to host the image, executing it within an Azure web app via Azure App Services \footnote{\url{https://executionengined4py.azurewebsites.net/run}}. This setup provides scalability and adaptability, creating an ideal environment for testing the \texttt{Laminar} framework.

We conducted latency tests for \texttt{Laminar} and compared them to regular \texttt{dispel4py} execution. The average execution time for each method and mapping was recorded in seconds. Notably, these tests involved direct execution without workflow registration.  However, it is important to note that for both local and remote \texttt{Laminar} experiments, the \textit{Registry} is hosted remotely on the web-based service (see Section~\ref{section:registry}).  

\begin{table}[h!]
\centering
\begin{tabular}{ |m{3.3em}|m{10em}|m{10em}|}
\hline
\textbf{Property} & \textbf{Local Ex. Engine} & \textbf{Remote Ex. Engine} \\
\hline
OS & Ubuntu 20.04 LTS & Unix 5.15.116.1 \\
\hline
Kernel & 5.10.16.3-microsoft-standard-WSL2 & N/A \\
\hline
CPU & Intel(R) Core(TM) i5-7200U CPU @ 2.50GHz & 
Intel(R) Xeon(R) CPU E5-2673 v4 @ 2.30GHz \\
\hline
Memory & 6.1Gi & 1.6Gi \\
\hline
\end{tabular}
\caption{Execution Engines Configuration}
\label{table:execution-engines}
\vspace{-0.2cm}
\end{table}

\begin{table}[h!]
    \centering
    \begin{tabular}{|c|c|c|}
        \hline \multirow{3}{*}{ Execution Method} & \multicolumn{2}{c|}{ Mapping } \\
        \cline{2-3} & \textit{Simple} & \textit{Multi} \\
        \hline \textit{original dispel4py}& 642 sec. & 7.32 sec. \\
        \hline \textit{Local Execution} (with \texttt{Laminar})& 928.2 sec. & 11.31 sec. \\
         \hline \textit{Remote Execution} (with \texttt{Laminar})& 1002 sec.  & 12.94 sec. \\
        \hline
    \end{tabular}
    \caption{Execution times of the \textit{Internal Extinction} }
    \label{table:laminar-performance-eval}
    \vspace{-0.3cm}
\end{table}

\vspace{-0.6cm }
The results in Table~\ref{table:laminar-performance-eval} reveal a noticeable latency between the original \texttt{dispel4py} execution (performed within the same computing environment as the one used for \texttt{Laminar}'s Local Execution Engine) and the utilization of the \texttt{Laminar} framework. \texttt{Laminar} engages in substantial processing, encompassing the retrieval of workflows and their associated PEs from the \textit{Registry}, their preparation for external execution, and an in-depth analysis to ensure dependency management. Additionally, the framework automatically installs the requisite Python libraries within the (local/remote) \textit{Execution Engine}. Furthermore, the request traverses an additional layer at the server, contributing to the observed latency. 

Conversely, examining the latency between local and remote \textit{Execution Engines} reveals no substantial increase. This suggests that the framework can harness cloud technology to enhance its performance and usability on a broader scale, without introducing significant performance concerns.

%\begin{table}[h!]
%    \centering
%    \begin{tabular}{|c|c|c|}
%        \hline \multirow{3}{*}{ Execution Method} & \multicolumn{2}{c|}{ Mapping } \\
%        \cline{2-3} & Sequential & Multiprocessing \\
%        \hline \textit{dispel4py}& 16.8  & 13.8  \\
%        \hline \textit{Local Execution (with Laminar)}& 499.43 & 565.32 \\
%         \hline \textit{Remote Endpoint (with Laminar)}& 635.29  & 685.66 \\
%        \hline
%    \end{tabular}
%    \caption{IsPrime Workflow Latency Results}
%   \label{table:laminar-performance-eval}
%\end{table}

\subsection {Deep learning Models}

\subsubsection{Semantic Code Search}
\label{section:ev-semantic-search}

As discussed in Section~\ref{section:code-search}, our choice of the fine-tuned \textit{unixcoder-code-search} model for generating embeddings plays a fundamental role in both PE description and user text input for semantic code searches. This decision stems from the model's proven adeptness in capturing intricate code semantics.

To reinforce our selection, in this work we conducted additional evaluations comparing the fine-tuned \textit{unixcoder-code-search} model with its base version~\footnote{\url{https://huggingface.co/microsoft/unixcoder-base}}. Using the Mean Reciprocal Rank (\texttt{MRR}) metric, consistent with prior studies\cite{unixcoder}, we comprehensively assessed the model's performance. The detailed results of these evaluations are presented in Table~\ref{table:code-search-eval}, providing further evidence of the model's effectiveness in enhancing semantic code search capabilities for natural language queries, specifically in the context of zero-shot\footnote{Zero-shot refers to the capability of a model to perform a search tasks without being explicitly trained on the specific queries or examples involved in the search.} \textit{text-to-code} search.

\begin{table}[h!]
    \centering
    \begin{tabular}{|c|c|c|}
        \hline \multirow{3}{*}{ Model } & \multicolumn{2}{c|}{ Zero-shot Code Search } \\
        \cline{2-3} & CosQA & CSN \\
        \cline{2-3} & \multicolumn{2}{c|}{ MRR } \\
        \hline \textit{unixcoder-base}&43.1  &44.7  \\
        \hline \textit{unixcoder-code-search}& \textbf{58.8}  & \textbf{72.2} \\
        \hline
    \end{tabular}
    \caption{Results on zero-shot text-to-code search}
    \label{table:code-search-eval}
\end{table}

\vspace{-0.5cm}
The evaluation employed two distinct datasets. CoSQA (Code Search and Question Answering)\cite{cosqa} consists of 20,604 labeled pairs of natural language queries and codes, annotated by at least 3 human annotators. The CSN dataset\cite{guo2021graphcodebert} is derived from the broader CodeSearchNet dataset, with curated filtering of low-quality queries. The \texttt{MRR} values for both models validate the superior performance of the fine-tuned \textit{unixcoder-code-search} model.

\subsubsection{Code Completion}
\label{section:ev-code-completion}

The zero-shot clone detection evaluation by ReACC\cite{lu2022reacc} involved assessing a model's ability to retrieve similar code segments from a dataset using partial queries, yielding positive outcomes.  In this work we have explored different large models for code completion (as discussed in Section~\ref{section:code-completion}), expanding the assessment to cover a variety of candidates, including an \textit{Unixcoder} model fine-tuned by our team for clone detection (\textit{unixcoder-clone-detection})\footnote{\url{https://huggingface.co/Lazyhope/unixcoder-clone-detection}}, the fine-tuned \textit{unix-code-search} model, and two state-of-the-art text embedding models: \textit{BAAI/bge-large-en}\footnote{\url{https://huggingface.co/BAAI/bge-large-en}} and \textit{thenlper/gte-large}\footnote{\url{https://huggingface.co/thenlper/gte-large}}. Additionally, we evaluated the CodeBERT~\cite{codebert} and GraphCodeBERT~\cite{graphcodebert} models, tailored specifically for source code analysis and comprehension.

\begin{table}[h!]
    \centering
    \begin{tabular}{|c|c|c|}
        \hline
        \textbf{Model} & \textbf{MAP@100} & \textbf{Precision at 1} \\
        \hline
        \textit{CodeBERT} & 1.47 & 4.75 \\ \hline
        \textit{GraphCodeBERT} & 5.31 & 15.68 \\ \hline
        \textit{ReACC-retriever-py} & 9.60 & \textbf{27.04} \\ \hline
        \textit{thenlper/gte-large} & 1.9 & 7 \\ \hline
        \textit{BAAI/bge-large-en} & 8.17 & 20 \\ \hline
        \textit{unixcoder-clone-detection} & \textbf{10.4} & 17 \\ \hline
        \textit{unixcoder-code-search} & 8.53 & 22.84 \\
        \hline
        \end{tabular}
    \caption{Zero-shot clone detection evaluation results}
    \label{table:code-completion-eval}
\end{table}

\vspace{-0.5cm}

For this evaluation, we utilized the CodeNet Python dataset\cite{puri2021codenet}, comprising around 14 million code samples as solutions to diverse coding problems. Our analysis employed two key metrics: \texttt{MAP@100} (Mean Average Precision at 100) and \texttt{Precision at 1}. \texttt{MAP@100} computes the average precision of the top 100 retrieved items for each query and then calculates the mean across all queries. In contrast, \texttt{Precision at 1} gauges the model's accuracy in retrieving the most relevant item as the \textbf{top recommendation}. The detailed results of this evaluation are presented in Table~\ref{table:code-completion-eval}. Since our primary interest lies in code completion, we place greater importance on the values obtained for \texttt{Precision at 1}. This metric directly signifies the model's proficiency in retrieving the most similar code from the dataset. As a result, we have chosen the \textit{ReACC-retriever-py} model for code retrieval due to its robust Precision performance.

\section{Related Work}
\label{section:related-work}

Several frameworks in the field of Function-as-a-Service (FaaS) and serverless computing have paved the way for advancements similar to \texttt{Laminar}. We discuss some of the key related frameworks below:

\begin{itemize}

\item FuncX~\cite{Li_2022} specializes in stateless function executions within distributed computing environments. However, it lacks the ability to handle streaming data and support stateful computations, distinguishing it from \texttt{Laminar}'s stream-based processing capabilities. 

\item PyWren~\cite{JonasVSR17} is an open-source FaaS platform for distributing Python functions over the cloud. Like FuncX, it supports stateless function execution and distributed computing, but lacks integrated streaming data management and deep learning code search features proposed in our work.  %which sets Laminar apart with its unique serverless stream-based processing and integrated deep learning search functionalities.

\item Apache OpenWhisk~\footnote{\url{https://openwhisk.apache.org/documentation.html}} is an open-source FaaS platform known for its flexibility and scalability in event-driven function execution. However, it does not inherently cater to the unique demands of streaming data processing and deep learning code search as does \texttt{Laminar}.

\item Apache Flink~\cite{katsifodimos2015apache} excels in distributed stream processing with low latency and high throughput, but lacks \texttt{Laminar}'s integrated serverless approach.

\item OpenFaaS~\footnote{\url{https://docs.openfaas.com/}} deploys event-driven functions using Docker and Kubernetes, yet lacks \texttt{Laminar}'s comprehensive stream processing and deep learning code search capabilities.

%\item Amazon Kinesis[REF], is an AWS service that  offers a scalable platform for real-time data streaming and analytics, handling diverse data sources efficiently. [NOTE FOR Rosa: DEVELOP MORE]

\end {itemize}

%\texttt{Laminar} stands out from these related frameworks by integrating stream-based processing and deep learning code search in a serverless environment. %It excels in handling streaming data and supporting stateful computations with seamless efficiency.

\section{Conclusions and Future Work}
\label{section:conclusions}

We present \texttt{Laminar}, a novel serverless stream-based framework with integrated deep learning search features. Unlike conventional serverless platforms, \texttt{Laminar} efficiently manages streaming data, accommodates stateless and stateful computations, and automatically parallelizes streaming applications across various enactment engines. Leveraging the \texttt{dispel4py} library, \texttt{Laminar} surpasses existing frameworks by providing a distinct approach to  streaming workflow and PE management, complete with automatic library detection and installation. This relieves users from manual installation and management of the execution engine's dependencies.

Moreover, \texttt{Laminar} integrates advanced large language models, introducing powerful deep learning code search and completion capabilities. This integration enables sophisticated PE code search through both code and natural language, facilitating seamless exploration and utilization of an extensive repository of PE functionalities. The framework also incorporates code summarization, automating PE functionality summaries and enhancing semantic search. Moving forward, we aim to integrate \texttt{Laminar} with popular cloud providers, enable multiple \textit{Execution Engine} registration, explore dynamic resource provisioning and container management strategies and enhance deep learning search for workflows.

\bibliographystyle{ACM-Reference-Format}
\bibliography{laminar-base}

\end{document}